\documentclass[apex]{jjap3}
\usepackage{lipsum}
\usepackage{graphicx}
\usepackage{newtxtext,newtxmath}
\usepackage{color}
\usepackage{graphicx,amsmath,amssymb,bm}
\usepackage{mathtools}
\usepackage{breqn}
\usepackage{ulem}

\title{Electronic Origin of Phase Stability in Mg--Zn--Y Alloys with a Long-Period Stacking Order}
\author{Takao Tsumuraya$^{1,2}$\thanks{E-mail: tsumu@kumamoto-u.ac.jp}, Hiroyoshi Momida$^{3}$, and Tamio Oguchi$^{4}$}
\inst{$^{1}$Priority Organization for Innovation and Excellence, Kumamoto University, Kumamoto 860-8555, Japan \\
$^{2}$Magnesium Research Center, Kumamoto University, Kumamoto 860-8555, Japan \\
$^{3}$The Institute of Scientific and Industrial Research, Osaka University, Ibaraki 567-0047, Japan\\
$^{4}$Center for Spintronics Research Network, Osaka University, Toyonaka 560-8531, Japan
}
\abst{
The origin of the phase stability of 18$R$ Mg--Zn--Y alloys with a long-period stacking order (LPSO) is studied using first-principles calculations. 
We calculate the heat of formation as a function of the number of Zn vacancies to discuss the role of Zn atoms.
The calculated convex hull indicates that the Zn atoms in the LPSO alloys are stable even if they number about half of the Y atoms.
The bonding state with Zn $p$ orbitals leads to the stability of the LPSO structure because the partial density of states of Mg nearest to the solute cluster forms a valley structure.
}
\begin{document}
\maketitle
Long-period stacking order (LPSO) is a unique crystal structure, where its stacking faults (i.e., Shockley partial dislocations) are ordered periodically along with the stacking direction of close-packed planes~\cite{Sato_CuAu1961, Lysak_1963, Oka_1972}.  
In the early 1950s, Suzuki proposed that when solute atoms enter a face-centered cubic (fcc) metal, the chemical potential of the solute element is different from that of the fcc metal matrix, resulting in segregation of the solute element from the stacking fault~\cite{HSuzuki1952,HSuzuki_JPSJ1962}. This phenomenon was observed in various alloy systems.~\cite{Kamino_1992, Han_Acta2003} 

In 2001, a dual phase magnesium (Mg) alloy with a nominal composition of Mg--1Zn--2Y (at.\%) composed of a LPSO structured phase and an $\alpha$-Mg phase was discovered to exhibit a tensile yield strength of more than 600MPa and an elongation to fracture of 5\%.~\cite{kawamura2001rapidly} 
Since then, attention has been focused on developing high-strength Mg-based alloys containing LPSO, which promotes kink strengthening mechanisms.~\cite{Hagihara2010, Shao_Kink2010} 

\begin{figure}[htpb]
\begin{center}
\includegraphics[width=0.85\linewidth]{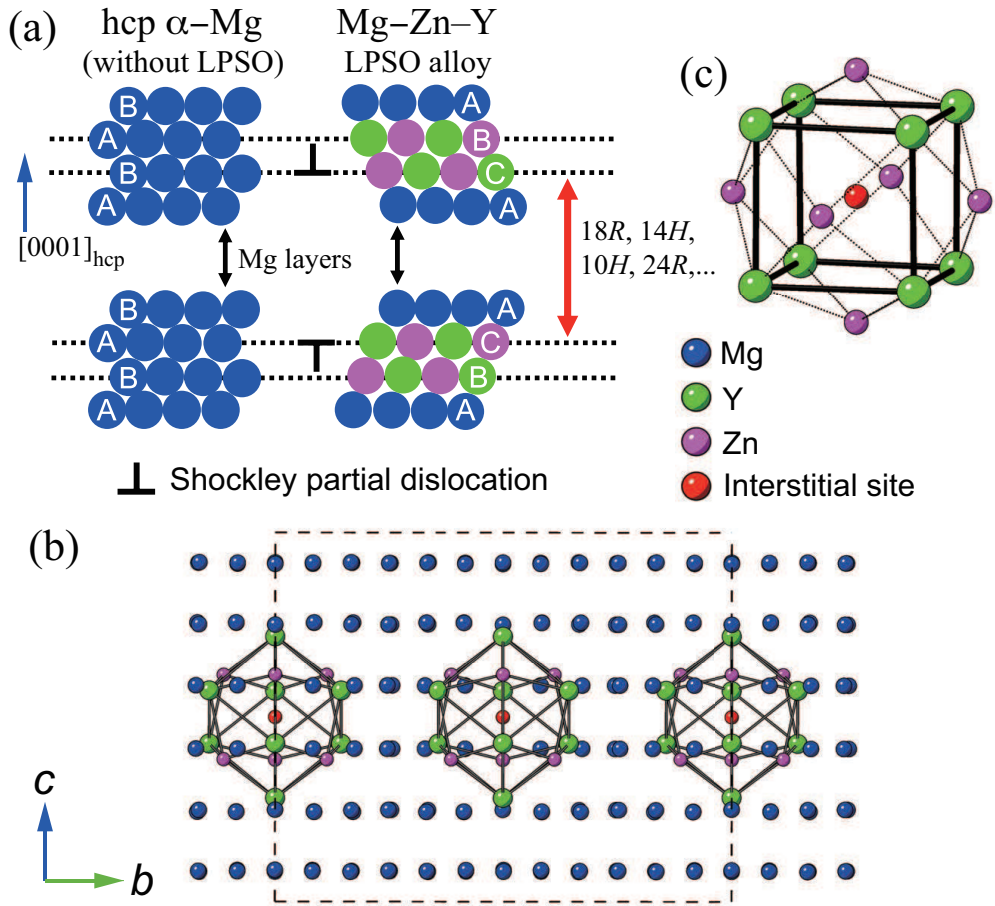}
\end{center}
\caption{(Color online)~(a) Schematic diagrams of the hcp $\alpha$-Mg and LPSO structure in Mg--Zn--Y alloy.
(b) LPSO structure of Mg$_{58}$Zn$_6$Y$_8$ with an interstitial atom (monoclinic structure with the $C$2/$m$ space group). 
The unit cell is depicted with the dashed line.
(c) Isolated Zn$_6$Y$_8$ $L$1$_2$-arranged cluster, 
where the Zn atoms move away from a plane formed by four Y atoms (bold lines). 
The cluster is rotated from (b) to clearly display the $L$1$_2$-arrangement of Zn and Y atoms. 
}
\setlength\abovecaptionskip{-4pt}
\label{Cryst}
\end{figure}

The LPSO structure in the Mg--Zn--Y alloys exhibits both chemical and stacking order. 
As illustrated in Fig.~\ref{Cryst}(a), a concentration of solute atoms (Zn and Y) on the (0001) plane appears in a few layers of the hexagonal close-packed (hcp) Mg matrix~\cite{abe2002long}.
As proposed by Suzuki,~\cite{HSuzuki1952} Shockley partial dislocations occur in the solute-enriched layers.
In early studies, this alloy was considered to have a 6$H_2$ structure with an ABCBCB$^{\prime}$ stacking sequence, where A and B$^{\prime}$ layers are enriched by Zn and Y.~\cite{ping2002local,abe2002long} 
Some later studies showed that the 6$H_2$ structure was incorrect; it is actually the 18$R$ structure.~\cite{Itoi_Scripta2004, MMatsuda_MSEA2005}
Three other structural polytypes of 10$H$, 14$H$, and 24$R$ were also identified using a scanning transmission electron microscope (STEM).~\cite{MMatsuda_MSEA2005, Zhu_Acta_2010, egusa_2012Acta, yamasaki10H, Kiguchi_MatTran2013} 
The STEM measurements also revealed that solute elements are enriched as chemically ordered $L$1$_2$-type Zn$_6$Y$_8$ clusters in the Mg matrix.~\cite{egusa_2012Acta} 
The presence of the $L$1$_2$ clusters causes anisotropic plastic deformation in Mg--Zn--Y alloys. 
The LPSO structure might prevent dislocations from moving across the cluster-arranged layer, namely, non-basal slips and deformation twins, and limit the movement only along the direction parallel to the close-packed plane (basal plane).~\cite{Hagihara2010, Shao_Kink2010, KIM2015}
However, it is unclear what role the chemical bonds involving solute elements play in the mobility of dislocations. 
Furthermore, several experiments have reported various chemical compositions of Zn and Y for the LPSO structure,~\cite{Daria_MaterDes2019} such as Mg--2Zn--4Y,~\cite{abe2002long} Mg--3Zn--10Y,~\cite{ping2002local} Mg--4Zn--8Y~\cite{Chino_LPSO2004}, Mg--2Zn--8Y,~\cite{Zhu_Scripta2009} and Mg--6Zn--9Y alloys~\cite{egusa_2012Acta} (at.\%). 

However, the arrangement of Zn and Y in the diverging compositions remains unclear due to the difficulty of the experimental characterization of structural properties. 
Some previous works with first-principles calculations have been performed for fully crystalline LPSO alloys,\cite{Saal_scripta2012, Ma_DFT2013, Saal_Acta2014, kishida_Acta2015, Ma_Jalcom2017} without considering the effect of vacancies.
More recently, Y atoms were found to play a role in reducing energy barriers to forming a stacking fault,\cite{Kawano2020} but the role of Zn atoms has not been clarified.

This study aims to understand the electronic origin of the phase stability of Mg--Zn--Y alloys containing LPSO using first-principles calculations based on density functional theory (DFT). 
In particular, we focus on the role of Zn in LPSO alloys and discuss the effects of the formation of Zn vacancies. 

Generally, the density of states (DOS) of metals with an hcp structure is characterized by a deep valley around the Fermi energy ($E_F$).\cite{JYamashita_BeMg73, JYamashita_Book73, Paxton1997}
The depth of the valley in DOS represents the magnitude of splitting of the peaks in the occupied and unoccupied states,\cite{JYamashita_BeMg73} which is an energy separation between the bonding and antibonding orbitals.\cite{Slater1965}
Furthermore, by comparing electronic states between hcp Mg and hcp Be, Inoue and Yamashita suggested that the degree of energy splitting between the two peaks (the width of the deep valley or dip) appears as a difference in the enthalpy of formation between the metals.\cite{JYamashita_BeMg73, JYamashita_Book73}
Mackintosh and Andersen derived the force theorem:\cite{springford1979} the change in the total energy of an electron system can be calculated as the difference of appropriate sums of the eigenvalue.\cite{HEINE19801, Skriver1982FT}
Therefore, the phase stability analysis based on DOS near Fermi energy may be understood by the energy change based on the force theorem.~\cite{Fe_LPSO_RRes2021} 

The 18$R$ LPSO structure is often observed in the Mg--Zn--Y system\cite{Zhu2012_18R14H} and thus the present calculations are based on an 18$R$ structure with a base-centered monoclinic structure (space group: $C$2/$m$), determined by STEM measurements.~\cite{Zhu_Acta_2010, Yokobayashi2011, egusa_2012Acta}
The 18$R$ crystal structure observed after heat treatment at 500$^\circ$C for 300 hours has another type of monoclinic cell (space group: $C$2/$c$)~\cite{kishida_Acta2015}.
The latter structure has twice as many atoms as the former. 
However, we use the former structure in this study because there is little difference in the formation energy\cite{kishida_Acta2015} and electronic states between these two structures.

The observed solute atoms cluster embedded in the Mg matrix is distorted from the ideal $L$1$_2$ structure as illustrated in Fig.~\ref{Cryst}(b), where Zn atoms move away from the plane formed by four Y atoms~[Fig.~\ref{Cryst}(c)]. 
Previous DFT studies have verified the distortion of the cluster through structural optimization, which creates a sizable interstitial site at the body center of the $L$1$_2$ cluster.~\cite{egusa_LPSO2012}
The inclusion of an atom at the central site stabilizes the LPSO structure.~\cite{egusa_LPSO2012, Saal_scripta2012, Ma_DFT2013, Saal_Acta2014, kishida_Acta2015, Itakura_Acta2021}
The calculations verified that heat of formation is gained by about 20 to 24 meV compared to the absence of an interstitial atom.~\cite{SuppMat}
To discuss the stability of Zn vacancies along a line in the ternary Mg--Zn--Y phase diagram (Zn composition ratio), we use the case of Zn as the central element.~\cite{SuppMat} 

In the present calculations, one-electron Kohn--Sham equations are solved self-consistently using a pseudopotential technique with plane wave basis sets adopting the projected augmented-wave method~\cite{PAW1994} implemented in~{\sc Quantum ESPRESSO}~\cite{QE2017}.
The exchange-correlation functional is the generalized gradient approximation proposed by Perdew, Burke, and Ernzerhof~\cite{GGA_PBE}. 
The pseudopotentials in PSlibrary were used\cite{DALCORSO_PP} and were generated using \texttt{atomic} code (version 6.3) with a pseudization algorithm proposed by Troullier and Martins~\cite{TroullierMartins}.
The cutoff energies for plane waves and potential were set at 90 and 588 Ry, respectively. 
The dimensions of the $\bm{k}$-point mesh for the 18$R$ LPSO structure are 6 $\times$ 6 $\times$ 4 using the Gaussian smearing method during self-consistent loops, where the width of the smearing function is 0.02~Ry. 
The convergence threshold is 10$^{-8}$~Ry.
The DOS calculations use the $\bm{k}$-point dimensions of 8 $\times$ 8 $\times$ 4 with the improved tetrahedron method.~\cite{Imptetra_Kawamura2014} 

\begin{figure}[htpb]
\begin{center}
\includegraphics[width=0.6\linewidth]{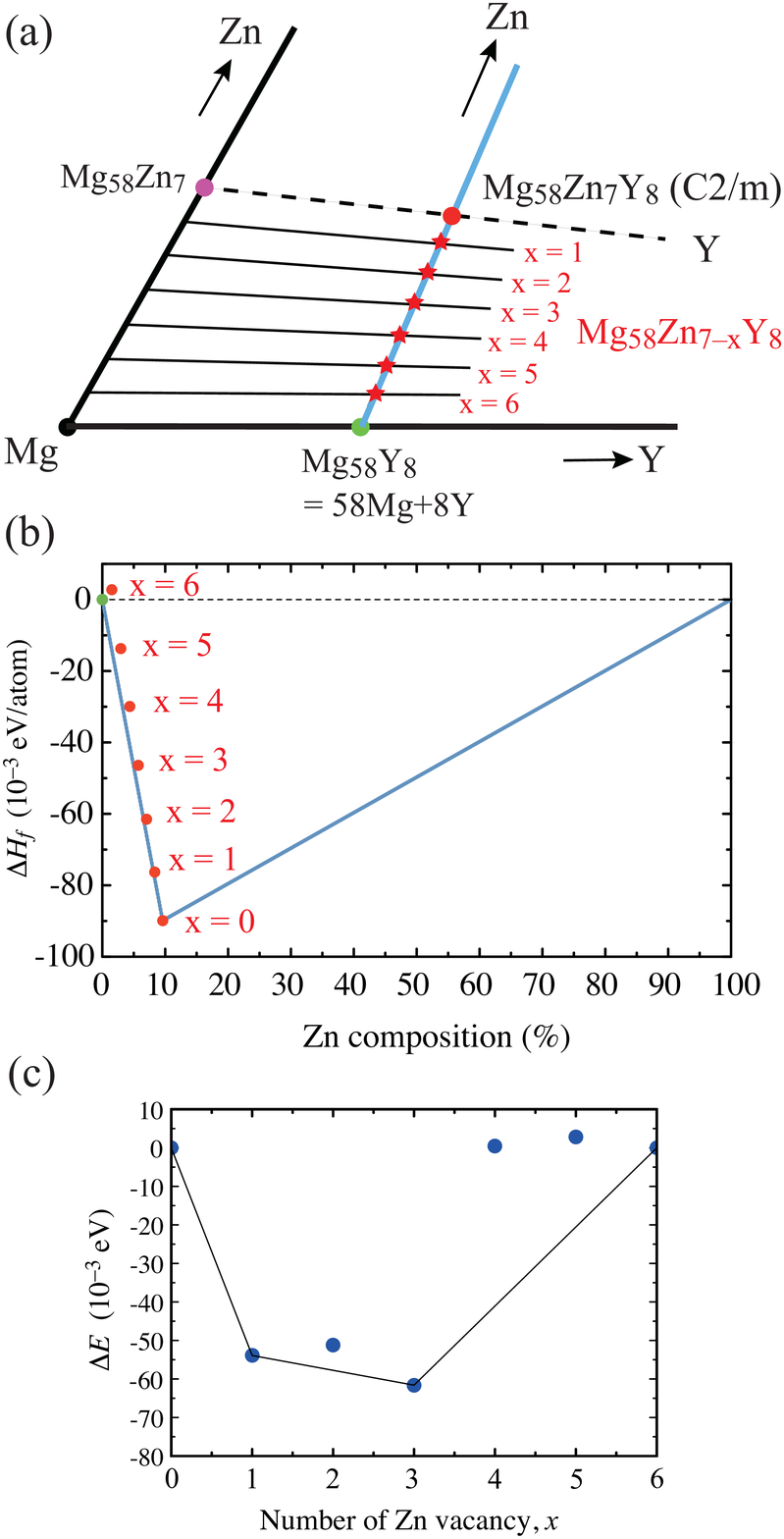}
\end{center}
\setlength\abovecaptionskip{-2pt}
\caption{
(Color online)~(a) Magnified section of ternary Mg--Zn--Y phase diagram. 
The red symbol on the diagram marks the constitution of Mg$_{58}$Zn$_{7}$Y$_8$, which indicates the number of atoms in the primitive unit cell of the 18$R$ structure. 
The red asterisk indicates the 18$R$ structure, including Zn vacancies, of Mg$_{58}$Zn$_{7-x}$Y$_8$, in which the number of Zn vacancies ($x$) is changed.
Mg$_{58}$Y$_8$ and Mg$_{58}$Zn$_7$ are pseudo compounds. 
(b) Scatter plot of $\Delta H_f$($x$) calculated based on Eq.~(\ref{eq1}).
The horizontal axis represents the Zn composition ratio (7$-$$x$)/(73$-$$x$).
We set the zero energy on the vertical axis at 58$E$(Mg) + 8$E$(Y) + (7$-$$x$)$E$(Zn) as the reference.
(c) Convex hull for the number of Zn vacancies ($x$) in Mg$_{58}$Zn$_{7-x}$Y$_8$ alloy, with reference energies set at $x$ = 0 and $x$ = 6.
The formation energy is calculated based on Eq.~(\ref{eq2}). 
}
\label{PhaseDiag}
\end{figure}

We first discuss the thermodynamic stability of the Zn vacancy by calculating the heat of formation as a function of the number of Zn vacancies ($x$). 
Figure~\ref{PhaseDiag}(a) presents a magnified section of the ternary Mg--Zn--Y phase diagram. 
The red filled circle in the phase diagram indicates the number of atoms in the primitive cell of the $C$2/$m$ (18$R$) structure with the interstitial atom of Zn, Mg$_{58}$Zn$_{7}$Y$_8$, and thus the total number of atoms is 73. 
The asterisks on the bold line indicate each structure of 18$R$-Mg$_{58}$Zn$_{7-x}$Y$_8$, including Zn vacancies from $x$ = 0 to 6, that is, the number of Zn atoms in Mg$_{58}$Zn$_{7-x}$Y$_8$ is deduced using the fixed number of Mg and Y atoms.

To understand whether the Mg--Zn--Y alloy, including Zn vacancies, can be formed from the elemental metals of Mg, Zn, and Y, 
the heat of formation for a specific Zn vacancy number, $x$, is defined as
\begin{align}
\label{eq1}
\begin{multlined}[b][4.5cm]
\Delta H_f (x) = \frac{1}{(73-x)} [E (\mathrm{Mg}_{58}\mathrm{Zn}_{7-x}\mathrm{Y}_8) \\ 
                 -{58 E(\mathrm{Mg}) + ({7-x}) E(\mathrm{Zn}) + 8 E(\mathrm{Y})}],
\end{multlined}
\end{align}
where $E$($\mathrm{Mg_{58}Zn_{7-x}Y_8}$) is the total equilibrium energy of the 18$R$-LPSO alloy with vacancy ($x$). 
For each vacancy state, we deduced all the possibilities of extracting $x$ atoms from the seven Zn sites in the Mg--Zn--Y alloy system. 
Subsequently, the lattice parameters and internal coordinates were optimized. 
The lattice parameters for $x$ = 0 are $a$ = 11.1, $b$ = 19.3, $c$ = 16.1 \AA, and $\beta$ = 76.6$^\circ$, which agree well with experiments [$a$ = 11.1, $b$ = 19.4, $c$ = 15.6 (= 46.89/3)~\AA, and $\beta$ = 83.3$^\circ$].~\cite{Zhu_Acta_2010, Zhu2012_18R14H}

In Figs.~\ref{PhaseDiag}(b) and~\ref{PhaseDiag}(c), we plot only the energy values for the most stable structure.
$E$($M$) denotes the equilibrium energy per atom of the elemental metals and is used as a reference of the chemical potential.
Details are presented in the Supplementary Material.~\cite{SuppMat}

Figure~\ref{PhaseDiag}(b)~displays the scatter plot of the heat of formation, $\Delta H_f$, as a function of the Zn composition ratio, calculated with Eq.~(\ref{eq1}). 
The Zn composition ratio (at\%) is calculated according to (7--$x$)/(73--$x$) $\times$ 100\%, which corresponds to the bold line in Fig.~\ref{PhaseDiag}(a). 
If the composition of Zn is zero ($x$ = 7), it is a pseudo compound of Mg$_{58}$Y$_8$.
The $\Delta H_f$ of Mg$_{58}$Y$_8$ calculated from the LPSO structure is unstable compared to the sum of the total energy of the elemental metals of hcp Mg and hcp Y.
Thus, in Fig.~\ref{PhaseDiag}(b), the energy references are set at the sum of the total energy of the elemental metals, 58$E$(Mg) + 8$E$(Y) + (7$-$$x$)$E$(Zn).  
Figure~\ref{PhaseDiag}(b) suggests that all phases, including Zn vacancies ($x$ = 1 $\sim$ 6), are slightly above the bold line connecting $x$ = 0 and $x$ = 7. 
Therefore, these phases are thermodynamically unstable and decompose into Mg$_{58}$Zn$_7$Y$_8$ ($x$ = 0) and elemental 58Mg+8Y ($x$ = 7).
Nevertheless, this diagram indicates that the distance from the hull decreases as the number of vacancies ($x$) decreases, where the energy differences between the red circles and the line connecting $\Delta H_f$ of $x$ = 0 and $x$ = 7 are closer.

Next, we consider the state in which Zn$_7$Y$_8$ clusters with an $L$1$_2$ structure are already formed in the Mg matrix and examine the convex hull when the ratio of Mg and Y is constant and only the atomic weight of Zn changes. 
For the convex hull, the formation energy as a function of $x$ is calculated as
\begin{align}
\begin{multlined}[b][4.5cm]
\Delta E(x) = E(\mathrm{Mg_{58}Zn_{7-x}Y_8}) \\
-\left\{\frac{(6-x)}{6} E(\mathrm{Mg_{58}Zn_7Y_8}) + \frac{x}{6} E(\mathrm{Mg_{58}ZnY_8})\right\}.
\end{multlined}
\label{eq2}
\end{align}
As plotted in Fig.~\ref{PhaseDiag}(c), the energies of $x$ = 1 and 3 lie on the convex hull. 
Therefore, incomplete $L$1$_2$ clusters with $x$ = 1 and 3 are stable, suggesting that this can be observed below the decomposition temperature. Especially, at $x$ = 3, the energy is the lowest in the convex hull and such an LPSO structure with a concentration ratio of Zn: Y = 1:2 is abundantly observed.~\cite{kawamura2001rapidly, abe2002long, Chino_LPSO2004, Itoi_Scripta2004}

On the contrary, the energy at $x$ = 2 is slightly above the line connecting $x$ = 1 and $x$ = 3, and the system decomposes into $x$ = 1 and $x$ = 3. 
However, $x$ = 2 is metastable since the energy difference between the energy at $x$ = 2 and the line connecting $x$ = 1 and 3 is relatively small.
In addition, the energies at $x$ = 4, 5, and 6 are unstable.
Therefore, the Zn atoms in the LPSO alloy are stable even if they are approximately half of the number of Y atoms.

The convex hull plotted in Fig.~\ref{PhaseDiag}(c) is obtained with the reference energy values at $x$ = 0 and $x$ = 6. 
However, the shape of the convex hull may change depending on the reference energies. 
The cases in which the reference energy is set at $x$ = 0 and $x$ = 4, and $x$ = 0 and $x$ = 5 are calculated and provided in the Supplementary Materials.~\cite{SuppMat} 
We came to the same conclusion as we did with $x$ = 0 and $x$ = 6 as the references. 

\begin{figure}[htb]
\begin{center}
\includegraphics[width=0.8\linewidth]{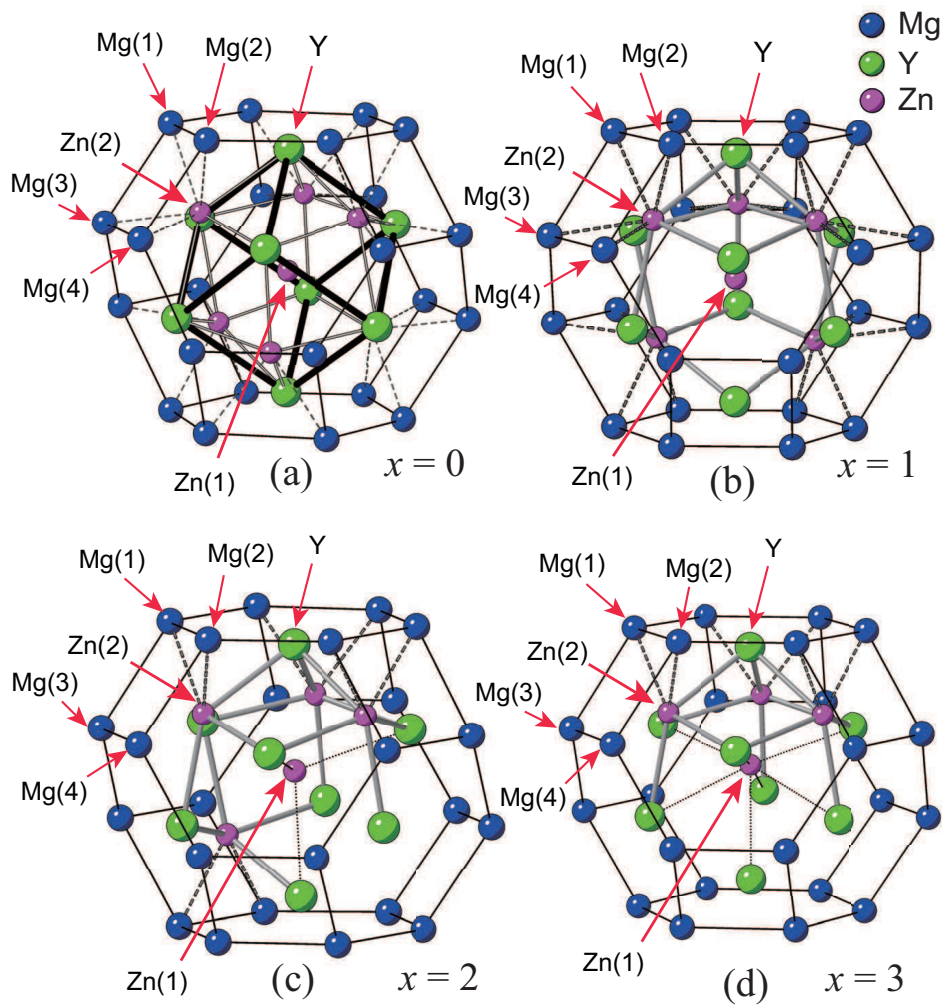}
\end{center}
\setlength\abovecaptionskip{-2pt}
\caption{(Color online)~Geometry of the solute atom cluster and surrounding Mg atoms in 18$R$ Mg$_{58}$Zn$_{7-x}$Y$_8$ alloy, where the number of Zn vacancies are (a) $x$ = 0, (b) $x$ = 1, (c) $x$ = 2, and (d) $x$ = 3. The most stable structure of each $x$ is presented.} 
\label{geometry}
\end{figure}

\begin{figure}[htb]
\begin{center}
\includegraphics[width=0.95\linewidth]{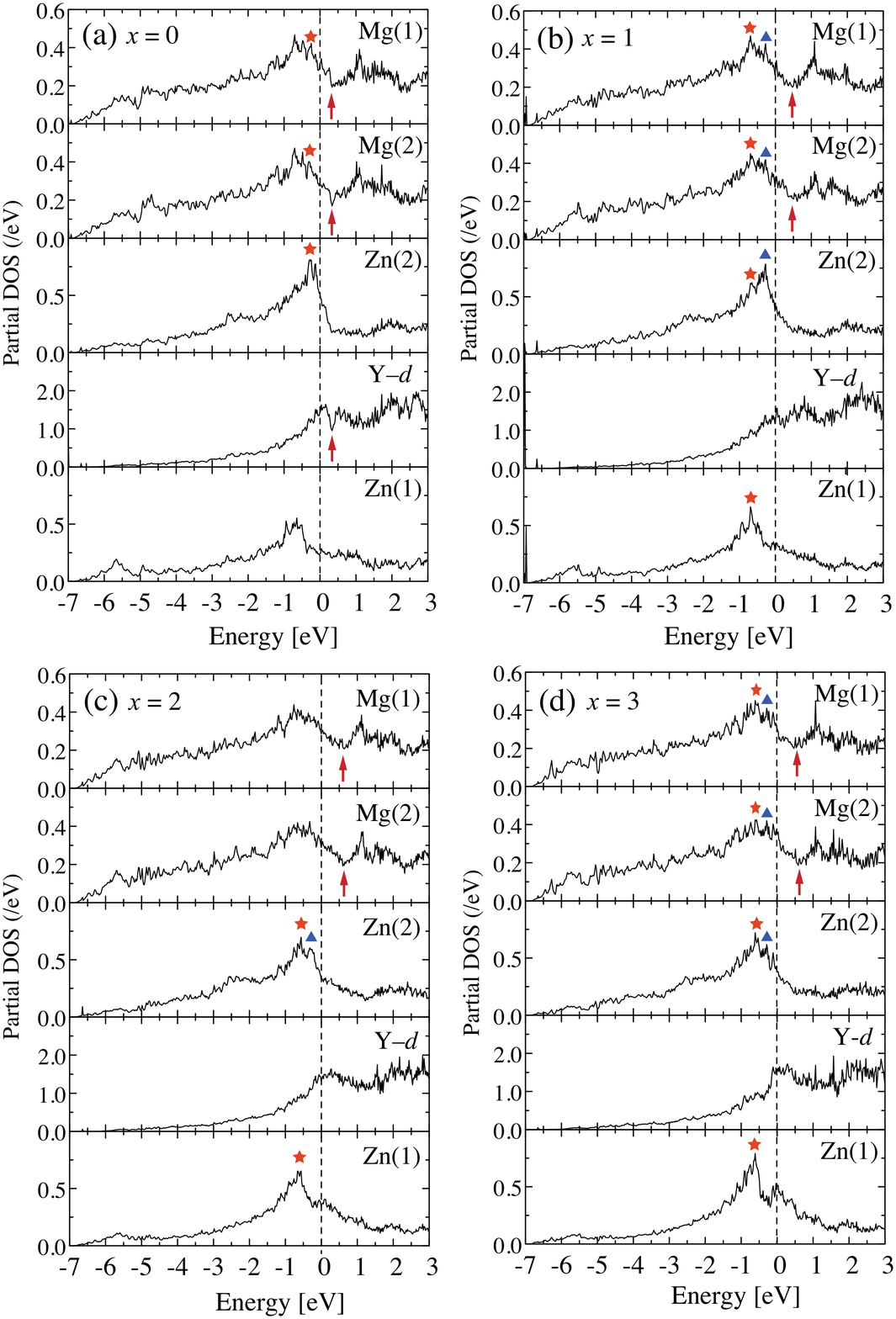}
\end{center}
\setlength\abovecaptionskip{-2pt}
\caption{
Partial density of states (DOS) projected on the $p$ orbitals of Mg, Zn, and $d$ orbitals of Y} in 18R-Mg$_{58}$Zn$_{7-x}$Y$_8$ alloy, where (a) $x$ = 0 (no vacancy), (b) $x$ = 1, (c) $x$ = 2, and (d) $x$ = 3.
The zero-energy with a longitudinal dashed line is the Fermi energy ($E_F)$. 
Up arrows represent the bottom of the DOS dip.
\label{PDOS}
\end{figure}
Finally, we examine the structural and electronic properties to elucidate the mechanism of stabilizing the Zn-vacancy-containing state. 
Figure~\ref{geometry} depicts the cluster of solute atoms and surrounding Mg atoms in the most stable structures of each $x$.
As depicted in Fig.~\ref{geometry}(a), the interstitial atom of Zn is called Zn(1), and a Zn atom that forms the $L$1$_2$ cluster is called Zn(2). 
Two pyramids share the Zn atoms forming Mg$_6$Y$_8$ cage: one square pyramid is formed with the square planar of Mg, and the other is included with Y. The four Mg atoms nearest to the Zn atom are referred to as Mg(1) -- Mg(4). 
In particular, the nearest-neighbor bonds to the Zn(2) atom are Mg atoms instead of Y atoms. 

Observing the distances between these four Mg atoms and the Zn atom in Fig.~\ref{geometry}, at $x$ = 1, the length between Zn(2) and Mg(2) decreased by 0.01~\AA~from that at $x$ = 0, due to the formation of the Zn vacancy.~\cite{SuppMat} 
At $x$ = 3, the distances from Zn(2) to both Mg(1) and Mg(2) decreased to 2.75~\AA,~compared to 2.78~\AA~at $x$ = 0.
These reductions in bond length indicate the strengthening of the bond between Mg and Zn.
However, the distances to Mg(3) and Mg(4) from Zn(2) do not change much.~\cite{SuppMat} 
Therefore, in Fig.~\ref{PDOS}, we only present the projected DOS on $p$ orbitals of Mg(1), Mg(2), Zn(1), Zn(2), and $d$ orbitals of Y in Mg$_{58}$Zn$_{7-x}$Y$_8$ alloy with $x$ = 0, 1, 2, and 3.

As mentioned earlier, the DOS of hcp metal has a valley and $E_F$ is located near the lowest position of the valley. 
The width of the valley appears as a difference in heat of formation between the metals.
The 18$R$ Mg--Zn--Y alloy also has a DOS valley.\cite{Ma_Jalcom2017} 
However, the lowest DOS position is approximately 0.2~eV higher than $E_F$,~\cite{Ma_Jalcom2017, SuppMat} as depicted in Fig.~\ref{PDOS}(a).
Therefore, its more energetically favorable phase could form by introducing electron deficiencies.

We first examined the dependence of $x$ on the total DOS.\cite{SuppMat}
It is discovered that the change in the DOS valley near $E_F$ was not distinct compared to simple metals with LPSO structure in Fe.~\cite{Fe_LPSO_RRes2021} 
Thus, we used the projected DOS of Zn and its neighboring Mg and plotted them in Fig.~\ref{PDOS}.
As shown by the asterisks and triangles in Figs.~\ref{PDOS}(b) and~\ref{PDOS}(d), the presence of Zn vacancies causes DOS peaks of $p$ states in Zn(2), Mg(1), and Mg(2) atoms around --0.8 eV. 
These Mg and Zn atoms form a bonding state below $E_F$, which reduces these partial DOS at $E_F$, and the width of the DOS valley of the Mg-$p$ states increases compared to the case without vacancies in Fig.~\ref{PDOS}(a).

The $p$ states of the Zn atoms also hybridize with the $d$ orbitals of the neighboring Y atoms, as seen in the DOS dip above the $E_F$ in Fig.~\ref{PDOS}(a).
The $d$ bands are partially filled with the $d^{1}$ configuration to form covalent bonds. 
The energy position (band filling) of the Y $d$ state is hardly changed when Zn vacancies are introduced [Figs.~\ref{PDOS}(b)--\ref{PDOS}(d)].
Therefore, Zn atoms in the LPSO alloy could be responsible for adhering Y atoms to the surrounding Mg atoms.

Furthermore, as $x$ increases, the partial DOS of Zn(1) at the interstitial sites becomes larger.
Especially at $x$ = 3, the $p$ states of Zn(1) hybridize with those of Zn(2), and the surrounding Mg atoms to form the bonding state below $E_F$, as shown in the peaks marked with asterisks in Fig.~\ref{PDOS}(d).

In conclusion, the convex hull we obtained at $x$ = 1 and 3, below the decomposition temperature, suggests that the Zn atoms in the LPSO alloy are stable even if they number about half of the Y atoms. 
Focusing on the partial DOS of Zn and its neighboring Mg (the boundary between the solute cluster and the Mg matrix), we reveal that the partial DOS of Mg atoms at the interface forms a valley structure due to the bonding state with Zn atoms.
The formation of the bonding state between Zn and the surrounding Mg atoms plays a crucial role in stabilizing the LPSO structure.

\bibliographystyle{jpsj}
\bibliography{./Mg_LPSO}
\acknowledgments
The authors thank Y. Kawamura, S. Ando, D. Shih, and M. Yamasaki for stimulating the discussion. This work was supported by the Cooperative Research Program of the Network Joint Research Center for Materials and Devices. This research was supported by a Grant-in-Aid for Scientific Research (Grants Nos. JP19K04988, JP20H00312, and JP22K04671) from the Japan Society for the Promotion of Science (JSPS) and a SICORP Grant (No.~JPMJSC2109) and a CREST Grant (No.~JPMJCR2094) from Japan Science and Technology Agency (JST). 
The calculations were conducted primarily at the Research Institute for Information Technology Computer Facilities at Kyushu University and MASAMUNE at the Institute for Materials Research, Tohoku University, Japan.  
\end{document}


\maketitle
\beginsupplement
\section{Effect of interstitial atoms on the electronic structure near the Fermi level}
~~We show how the interstitial atom in the Zn$_6$Y$_8$ $L$1$_2$ cluster affects the heat of formation and the total density of states (DOS) around the Fermi energy ($E_F$). 
Egusa $et$ $al$. originally reported that introducing an atom at the center of the Zn$_6$Y$_8$ cluster stabilizes the LPSO structure.~\cite{egusa_LPSO2012}
For the 18$R$ LPSO structure, we verified that the calculated heat of formation increased by approximately 20--25~meV/atom compared to that without an interstitial atom. 
The heat of formation for the 18$R$ Mg-Zn-Y alloys is calculated as
\begin{eqnarray}
    \Delta H &=& \left\{E (\mathrm{Mg_{58}Zn_6Y_8})-[58E(\mathrm{Mg}) + 6E(\mathrm{Zn}) + 8E(\mathrm{Y})]\right\}/72\\
             &=& -70~\rm{meV/atom}. \nonumber
\end{eqnarray}
where 72 is the number of atoms in the primitive cell of the $C$2/$m$ structure without an interstitial atom. 
$E$(Mg), $E$(Zn), and $E$(Y) are the equilibrium total energy of the elemental metals.

With an interstitial atom, the total number of atoms is 73, and the heat of formation is calculated as
\begin{eqnarray}
    \Delta H_{int}^{\rm{Mg}} &=& \left\{E (\mathrm{Mg_{59}Zn_6Y_8})-[59E(\mathrm{Mg}) + 6E(\mathrm{Zn}) + 8E(\mathrm{Y})]\right\}/73\\
                             &=& -95~\rm{meV/atom} \nonumber
\end{eqnarray}
\begin{eqnarray}
    \Delta H_{int}^{\rm{Zn}} &=& \left\{E (\mathrm{Mg_{58}Zn_7Y_8})-[58E(\mathrm{Mg}) + 7E(\mathrm{Zn}) + 8E(\mathrm{Y})]\right\}/73\\
                            &=& -90~\rm{meV/atom.} \nonumber
\end{eqnarray}
\begin{eqnarray}
    \Delta H_{int}^{\rm{Y}} &=& \left\{E (\mathrm{Mg_{59}Zn_6Y_9})-[58E(\mathrm{Mg}) + 6E(\mathrm{Zn}) + 9E(\mathrm{Y})]\right\}/73\\
                             &=& -94~\rm{meV/atom} \nonumber
\end{eqnarray}
The order of stability is Mg, Y, and Zn for the types of interstitial atoms. 
From Table~\ref{deltaH}, these results generally agree with those from previous studies.\cite{Saal_scripta2012, Saal_Acta2014, kishida_Acta2015, Itakura_Acta2021} 
The calculated heat of formation with a Zn interstitial atom ($-$90 meV/atom) is higher than that with Mg and Y atoms of $-$95 meV/atom and $-$94 meV/atom, respectively. However, Zn and Y have been experimentally reported to be in the central position of the cluster~\cite{kishida_Acta2015}.

\begin{table}[htpb]
\caption{Calculated heats of formation $\Delta$$H_f$ for 18$R$ LPSO structure with different types of interstitial atoms.}
\label{deltaH}
\begin{tabular}{clccc}
\Hline
Interstitials & Formula & $\Delta$$H_f$ (meV/atom)  &  Previous studies\\
\hline
No atom &  Mg$_{58}$Zn$_6$Y$_8$ & $-$70 & $-$70\cite{Saal_scripta2012}\\
Mg &  Mg$_{59}$Zn$_6$Y$_8$  & $-$95 & $-$98\cite{Saal_Acta2014}\\
Y  &  Mg$_{58}$Zn$_6$Y$_9$  & $-$94 & \\
Zn &  Mg$_{58}$Zn$_7$Y$_8$  & $-$90 & \\
\Hline
\end{tabular}
\end{table}

\begin{figure}[ht]
\begin{center}
\includegraphics[width=0.65\linewidth]{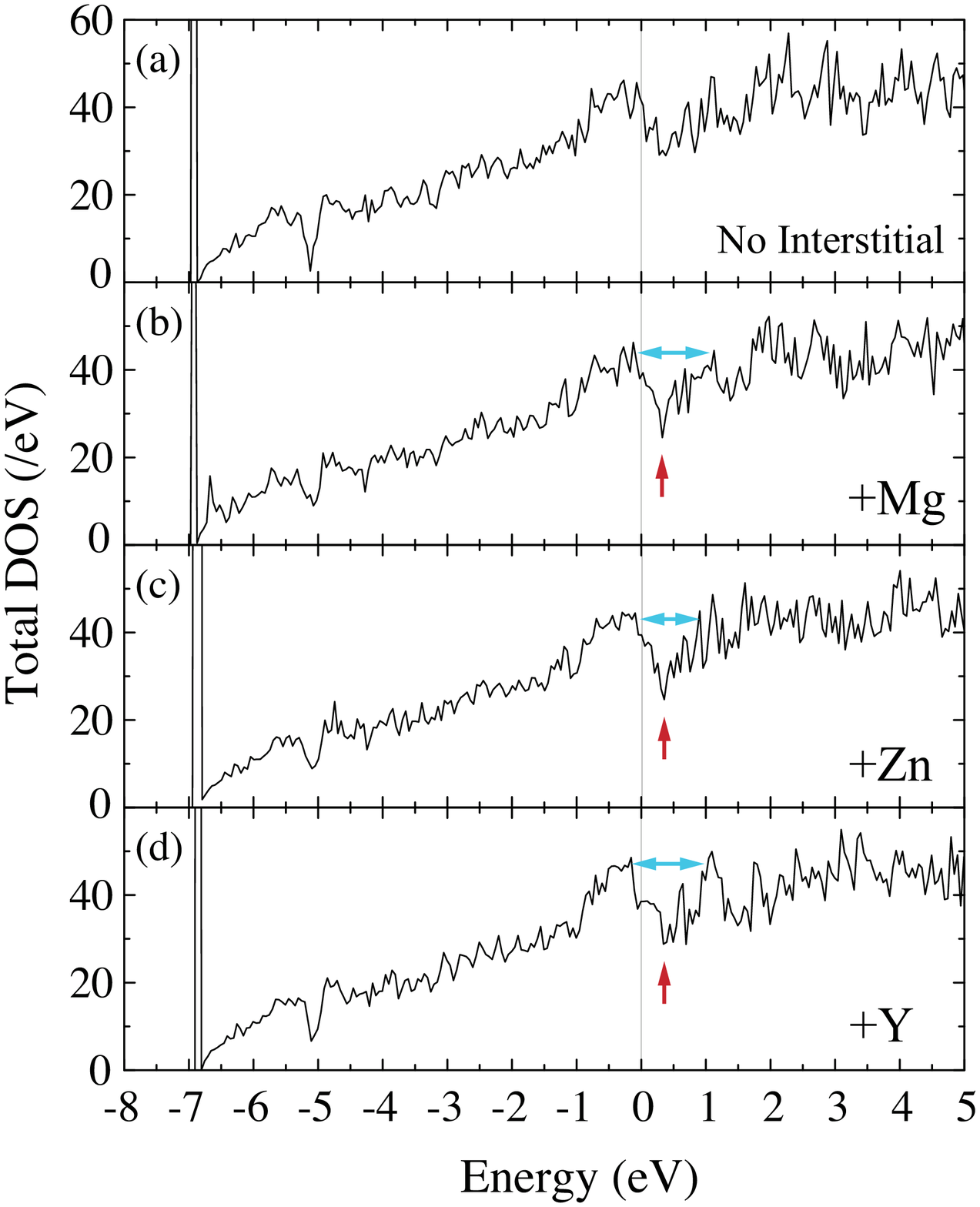}
\end{center}
\setlength\abovecaptionskip{-2pt}
\caption{
Total DOS of the 18$R$-type Mg-Zn-Y alloy (a) without an interstitial atom, 
and those with an interstitial atom of (b) Mg, (c) Zn, and (d) Y are depicted. 
The origin in the horizontal axis is taken at the Fermi energy~$E_F$. 
Double-headed arrows indicate the width of the DOS valley.
Up arrows represent the bottom of the DOS valley. 
}
\label{TDOS_interst}
\end{figure}

\begin{figure}[tb]
\begin{center}
\includegraphics[width=0.6\linewidth]{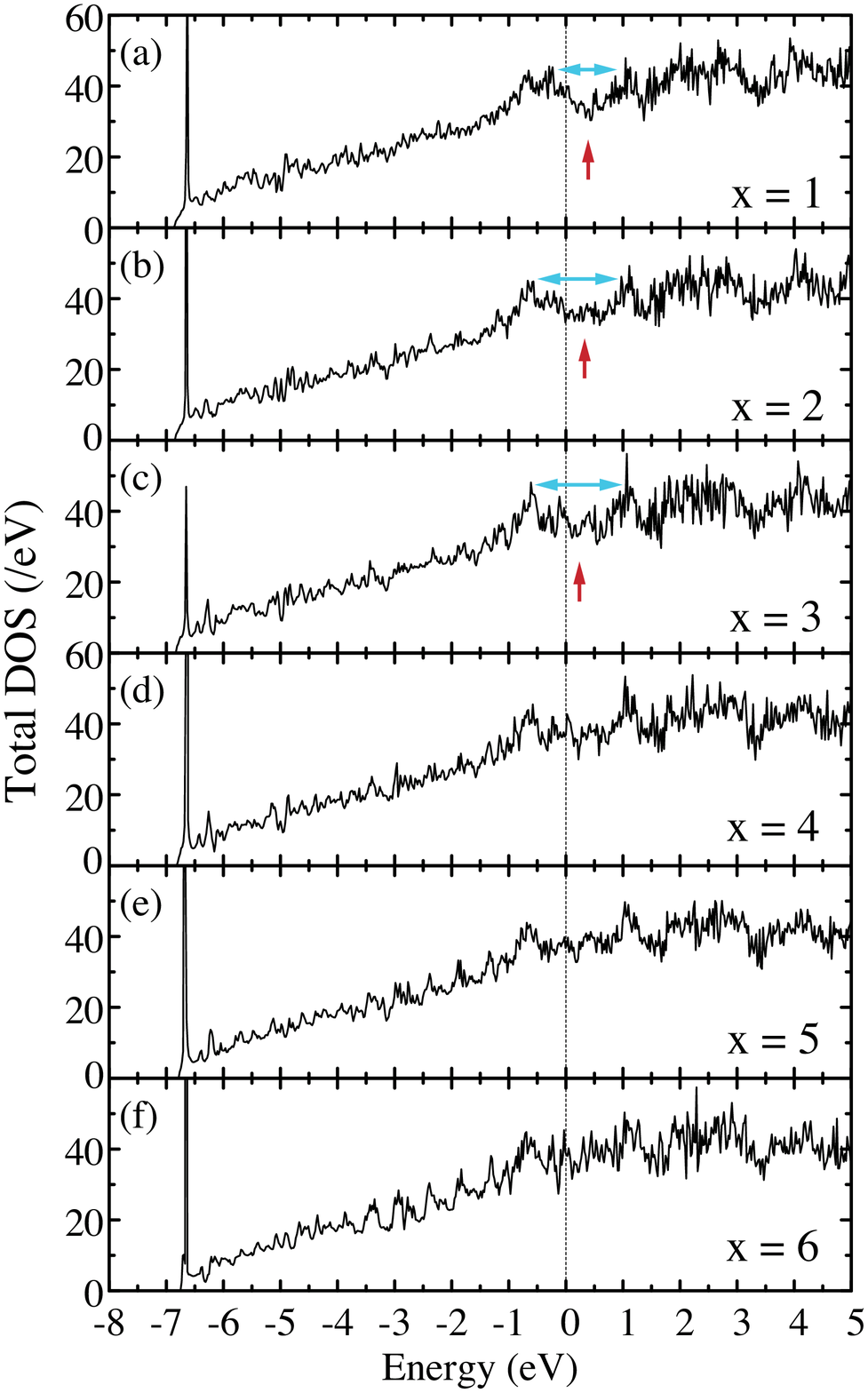}
\end{center}
\setlength\abovecaptionskip{-4pt}
\caption{Total DOS dependent on number of Zn vacancies $x$ in 18R-Mg$_{58}$Zn$_{7-x}$Y$_8$ alloy: (a) $x$ = 1; (b) $x$ = 2; (c) $x$ = 3; (d) $x$ = 4; (e) $x$ = 5; (f) $x$ = 6. Up arrows represent the bottom of the DOS valley.
Double-headed arrows indicate the width of DOS valley.
}
\label{tdos}
\end{figure}

We show that an increase in the heat of formation by adding an interstitial atom is related to the total DOS near the Fermi energy $E_F$.
As described in the main text, the DOS of the metal with a hexagonal close-packed (hcp) structure is generally characterized by the presence of a deep valley or dip near $E_F$, where $E_F$ is located at the lowest position of the DOS valley.\cite{Slater1965, JYamashita_BeMg73, JYamashita_Book73, Paxton1997} 
The depth of a DOS valley represents the magnitude of the splitting of the main DOS peaks between those in occupied and unoccupied states.
The width of a valley or dip near $E_F$ appears as a difference in the heat of formation (or cohesive energy) between different types of hcp metals.\cite{JYamashita_BeMg73, JYamashita_Book73}  
 
Figure~\ref{TDOS_interst}(a) plots the total DOS of the 18$R$-type Mg$_{58}$Zn$_6$Y$_8$ alloy in the absence of an interstitial atom. 
The DOS dip is observed, but is relatively shallow. The lowest position of the DOS is approximately 0.2 eV higher than the $E_F$. 
By introducing an Mg, Zn, or Y atom into the center of the cluster, the DOS valley near $E_F$ is more apparent, and the valley width increases compared to the case without an interstitial atom, as shown in Figs.~\ref{TDOS_interst}(b) -- \ref{TDOS_interst}(d),. 
Especially with an interstitial Y atom, the DOS at $E_F$ is reduced, and the width of the DOS valley increases. 

\section{Dependence of number of Zn vacancies~($x$) on total DOS}
~~The dependence of the number of Zn vacancies~$x$ on the total DOS is shown in Fig.~\ref{tdos}.
As $x$ increases from $x$ = 1 to $x$ = 3 in Mg$_{58}$Zn$_{7-x}$Y$_8$, 
the valley structure of DOS above the $E_F$ becomes shallow, and the valley width increases.
The Fermi level shifts to a high energy position close to the bottom of the DOS valley. 
The tendency of DOS also suggests that $x$ = 4 and $x$ = 5 are not stable. 
However, the change in total DOS in the LPSO-type Mg--Zn--Y alloys is not distinct compared to simple metals with LPSO structures in Fe.~\cite{Fe_LPSO_RRes2021} 
Thus, we used the projected DOS on the $p$ orbitals of the Zn and Mg atoms and the $d$ orbitals of the Y atom in the main text to clarify the chemical bond between them and the origin of structural stability. 

\newpage
\section{Convex hull diagram}
\begin{figure}[htpb]
\begin{center}
\includegraphics[width=0.6\linewidth]{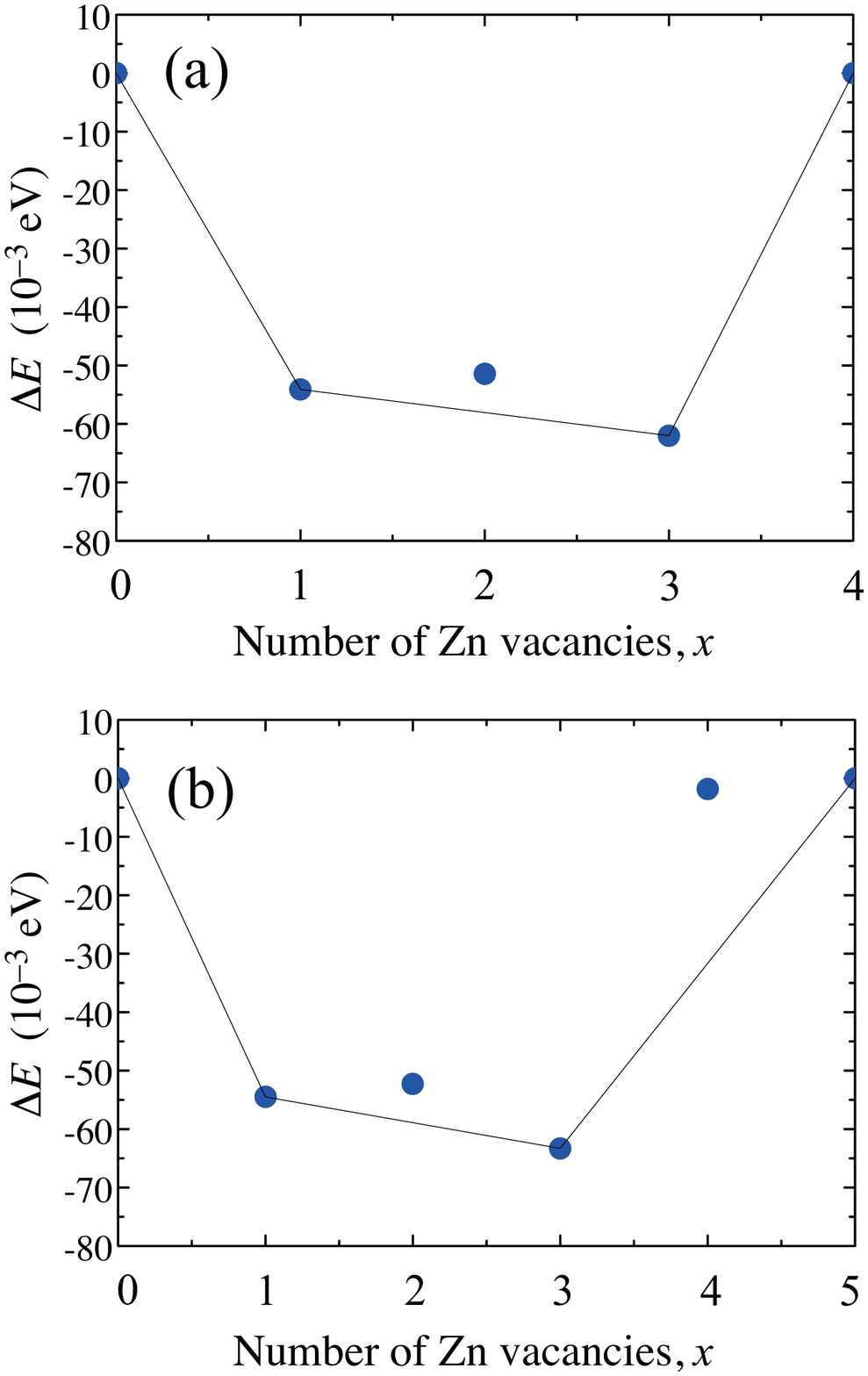}
\end{center}
\setlength\abovecaptionskip{-2pt}
\caption{
Convex hull diagrams for number of Zn vacancies ($x$) in 18$R$ LPSO structure with chemical formula of Mg$_{58}$Zn$_{7-x}$Y$_8$, 
with reference energies are set at (a) $x$ = 0 and $x$ = 4, and (b) $x$ = 0 and $x$ = 5.  
The formation energies are calculated using Eqs.~(\ref{eqS1}) and~(\ref{eqS2}), respectively. 
}
\label{Convex45}
\end{figure}

~~The convex hull plotted in Fig.~2(c) in the main text is obtained with reference energy values at $x$ = 0 and $x$ = 6. 
However, the shape of the convex hull may change depending on the choice of reference energy. 
The formation energy as a function of vacancies $x$ with reference energy values at $x$ = 0 and $x$ = 4 is calculated as
\begin{align}
\label{eqS1}
\begin{multlined}[b][4.5cm]
\Delta E(x) = E(\mathrm{Mg_{58}Zn_{7-x}Y_8})\\
-\left\{\frac{(4-x)}{4} E(\mathrm{Mg_{58}Zn_7Y_8}) + \frac{x}{4} E(\mathrm{Mg_{58}Zn_3Y_8})\right\},
\end{multlined}
\end{align}
and with reference energy values at $x$ = 0 and $x$ = 5 is calculated as
\begin{align}
\label{eqS2}
\begin{multlined}[b][4.5cm]
\Delta E(x) = E(\mathrm{Mg_{58}Zn_{7-x}Y_8})\\
-\left\{\frac{(5-x)}{5} E(\mathrm{Mg_{58}Zn_7Y_8}) + \frac{x}{5} E(\mathrm{Mg_{58}Zn_2Y_8})\right\}.
\end{multlined}
\end{align}
The convex hulls obtained with the reference energy values at $x$ = 0 and $x$ = 4, and $x$ = 0 and $x$ = 5 are plotted in Figs.~\ref{Convex45}(a) and \ref{Convex45}(b), respectively. 
We came to the same conclusion as we did with $x$ = 0 and $x$ = 6 as the references.

\section{Details of structural optimizations}
\subsection{Elemental hcp metal}
~~The lattice parameters of the hcp structure of elemental metals (Mg, Zn, and Y) are also optimized using the first-principles method.
The total energies are used as a reference to obtain the heat of formation. 
The dimensions of the mesh of $\bm{k}$ points for hcp metals is 20 $\times$ 20 $\times$ 8. 
We used the stress tensor for structural relaxations and set the cutoff energies for plane waves and potential as 90 Ry and 588 Ry, respectively. 
Table~\ref{hcpmetals} presents the DFT-optimized lattice parameters. 
The hcp Mg shows a nearly ideal $c$/$a$ ratio. Zn has a $c$/$a$ ratio greater than the ideal value of $\sqrt{8/3}$ = 1.633. 
In contrast, hcp Y has a smaller $c$/$a$ ratio than ideal.  
These results are in agreement with those reported in previous studies.~\cite{Albrecht2021} 

\begin{table}[htpb]
\caption{Optimized structural parameters for hcp structure of elemental metals. The ground states of these metals have an hcp structure with the space group of $P$6$_3$/$mmc$. 
The internal position is 2$d$ site at (1/3, 2/3, 3/4). }
\label{hcpmetals}
\begin{tabular}{ccccc}
\Hline
Element & $a$ & $c$ & $c$/$a$ &  \\
        & [\AA] & [\AA] &  &  \\
\hline
Mg &  3.19 & 5.17 & 1.619 & \\
Zn &  2.66 & 4.93 & 1.854 & \\ 
Y  &  3.65 & 5.73 & 1.572 & \\
\Hline
\end{tabular}
\end{table}

\subsection{18$R$ LPSO structure}
~~To obtain the calculated heats of formation, we performed structural optimization for the lattice parameters and internal atomic coordinates of the Mg-Zn-Y alloys containing Zn vacancies. 
The convergence threshold for forces for ionic minimization is 0.0001~Ry/Bohr. 
The calculated total energy has a precision of 0.001~Ry/atom; the difference between different structures (heat of formation) has a higher precision of approximately 0.0001~Ry/atom. 
We checked the convergence of $\bm{k}$-point meshes and the plane-wave cutoffs for the wavefunctions and charge densities. 
The cutoff energies for plane waves, and charge densities and potential are set at 90 and 588 Ry, respectively. 
The dimensions of the $\bm{k}$-point mesh are 6 $\times$ 6 $\times$ 4 using the Gaussian smearing method during self-consistent loops, where the width of the smearing function is 0.02~Ry. 
The convergence threshold for the self-consistent loop is 10$^{-8}$~Ry.

\begin{figure*}[htb]
\begin{center}
\includegraphics[width=0.98\linewidth]{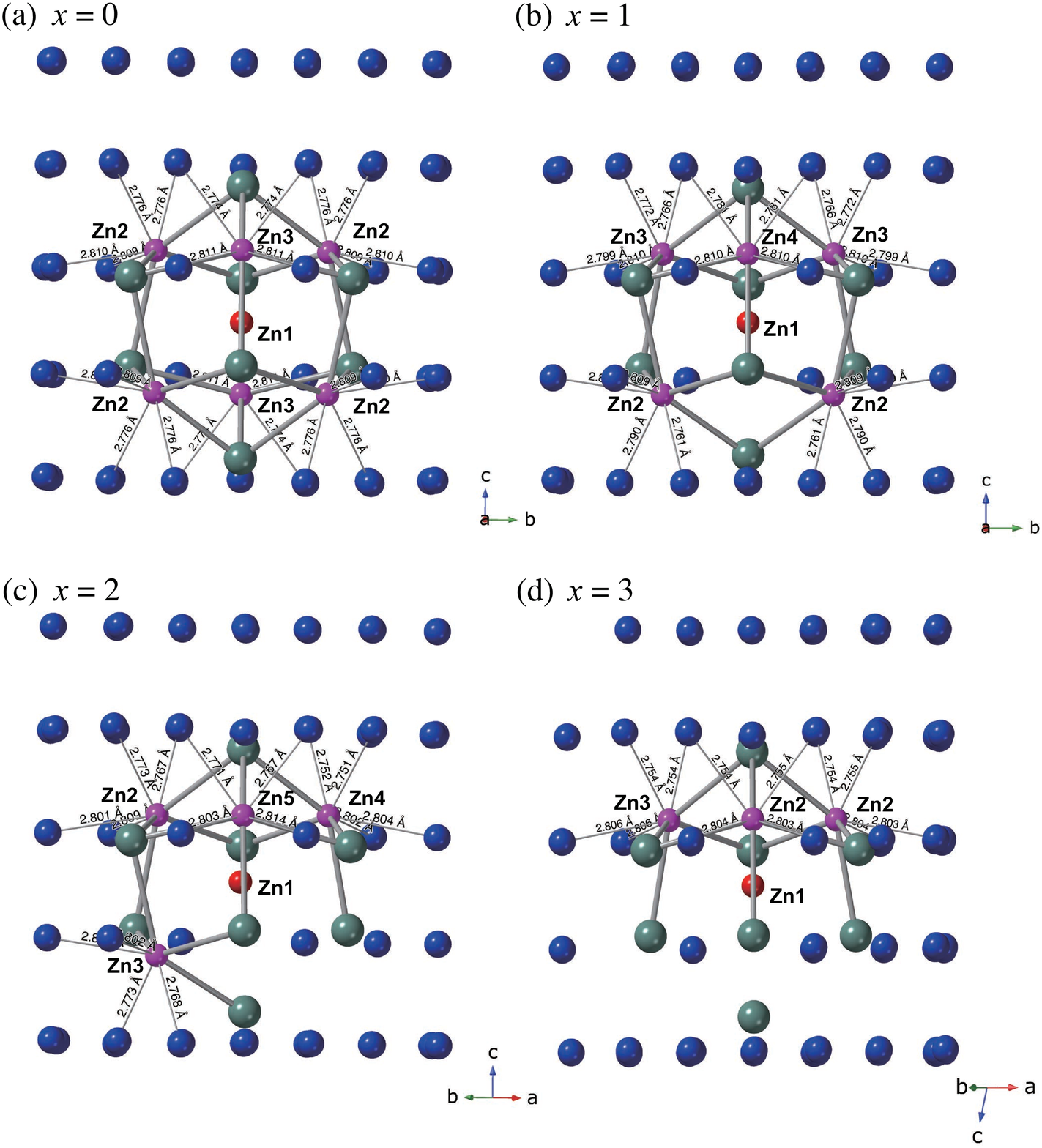}
\end{center}
\setlength\abovecaptionskip{-2pt}
\caption{Atomic structure near the deformed $L$1$_2$ cluster of solute atoms in 18$R$ Mg$_{58}$Zn$_{7-x}$Y$_8$ alloy, where (a) $x$ = 0, (b) $x$ = 1, (c) $x$ = 2, and (d) $x$ = 3. The distances between Zn and the surrounding Mg atoms are indicated. 
The blue, purple, and green spheres represent Mg, Zn, and Y, respectively. 
The $c$ axis in these figures is perpendicular to the basal (0001) plane of the hexagonal lattice. 
}
\label{FigdistanceMgZn}
\end{figure*}

Table~\ref{Zn2Mgdistance} presents the distances from the Zn atoms forming the cage composed of Zn and Y atoms to the four closest Mg atoms in the theoretically optimized structures. Their atomic geometries are depicted in Fig.~\ref{FigdistanceMgZn}.
Their atomic coordinates and lattice parameters are presented in Tables~\ref{Coordx0},~\ref{Coordx1},~and~\ref{Coordx3}, respectively.
Of the four bonds between the closest Mg atoms to the Zn atom, at $x$ = 1, 
the length between Zn2 (Zn3) and Mg15 (Mg16) decreased by 0.01~\AA~from that at $x$ = 0, due to the formation of a Zn vacancy.
At $x$ = 3, the distances from Zn2 (Zn3) to Mg15 (Mg20) and Mg13 (Mg20) decreased to~$\sim$2.755~\AA, compared to ~$\sim$2.775~\AA~at $x$ = 0.
These reductions in bond length suggest a strengthening of the bond between Mg and Zn.

\begin{table}[htpb]
\caption{Interatomic distances ($d$) to the four closest Mg atoms from Zn atom forming the Zn$_6$Y$_8$ cage in the theoretically optimized structure. The labels of the four nearest Mg atoms in the atomic coordinates are shown in parentheses. The atomic coordinates at $x$ = 0, 1, and 3 are presented in Tables~\ref{Coordx0}, \ref{Coordx1} and \ref{Coordx3}, respectively. 
The labels in square brackets correspond to the labels of atoms in Fig.3 (in the main text).  
}
\label{Zn2Mgdistance}
\begin{tabular}{|cccllll|}
\Hline
\Hline
$x$ = 0 & Label & Wyckoff position &  $d$(Zn--Mg)~[\AA] & $d$(Zn--Mg)~[\AA] & $d$(Zn--Mg)~[\AA] & $d$(Zn--Mg)~[\AA] \\
\Hline
& Zn2 & 8$j$ &  2.776 (Mg8) & 2.776 (Mg7) & 2.809 (Mg11) & 2.810 (Mg13) \\
Fig.3(a) & [Zn(2)] &    &  [Mg(1)] & [Mg(2)]  & [Mg(3)]  & [Mg(4)] \\
\hline
& Zn3 & 4$i$ &  2.774 (Mg10) & 2.774 (Mg10) & 2.811 (Mg12) & 2.811 (Mg12)\\ 
\Hline
$x$ = 1 & Zn2 & 4$b$ &  2.789 (Mg13) & 2.761 (Mg15) & 2.810 (Mg25) & 2.809 (Mg21)\\
\hline
& Zn3 & 4$b$ &  2.772 (Mg14) & 2.766 (Mg16) & 2.810 (Mg22) & 2.799 (Mg26)\\ 
Fig.3(b) & [Zn(2)] &    &  [Mg(1)] & [Mg(2)]  & [Mg(3)]  & [Mg(4)]  \\
\hline
& Zn4 & 2$a$ & 2.781 (Mg19) & 2.781 (Mg19)  & 2.810 (Mg24) & 2.810 (Mg24) \\ 
\Hline
$x$ = 2 & Zn2 & 1$a$ &  2.767 (Mg31) & 2.773 (Mg26) & 2.809 (Mg43) & 2.801 (Mg50)\\
\hline
& Zn3 & 1$a$ &  2.768 (Mg29) & 2.773 (Mg27) & 2.802 (Mg41) & 2.804 (Mg51)\\ 
\hline
& Zn4 & 1$a$ &  2.751 (Mg28) & 2.752 (Mg30) & 2.804 (Mg52) & 2.802 (Mg42)\\ 
\hline
& Zn5 & 1$a$ &  2.771 (Mg37) & 2.767 (Mg39) & 2.803 (Mg47) & 2.814 (Mg46)\\ 
Fig.3(c) & [Zn(2)] &    &  [Mg(1)] & [Mg(2)]  & [Mg(3)]  & [Mg(4)]  \\
\Hline
$x$ = 3 & Zn2 & 4$b$ &  2.755 (Mg15) & 2.754 (Mg13) & 2.803 (Mg21) & 2.804 (Mg25) \\
\hline
& Zn3 & 2$a$ &  2.754 (Mg20) & 2.754 (Mg20) & 2.806 (Mg23) & 2.806 (Mg23)\\ 
Fig.3(d) & [Zn(2)] &    &  [Mg(1)] & [Mg(2)]  & [Mg(3)]  & [Mg(4)]  \\
\Hline
\Hline
\end{tabular}
\end{table}

\begin{table}[htpb]
\setlength\abovecaptionskip{-2pt}
\caption{
Optimized atomic positions for 18$R$ Mg$_{58}$Zn$_7$Y$_8$ without Zn vacancy ($x$ = 0).
The space group is $C2/m$ (No.~12) and the optimized lattice parameters are  $a$ = 11.16, $b$ = 19.34, $c$ = 16.07 \AA, and $\beta$ = 103.48$^\circ$.
}
\label{Coordx0}
\small
\begin{tabular}{lccccc}
\Hline
      &  &  & fractional coordinates &  \\
      \hline
Label & Wyckoff position & $x$ & $y$ & $z$ \\
\Hline
Mg1 & 8$j$ & --0.44129 & 0.41783 & --0.08188 \\
Mg2 & 8$j$ & --0.44602 & 0.25131 & --0.08269 \\
Mg3 & 8$j$ & --0.44458 & 0.08215 & --0.08439 \\
Mg4 & 8$j$ & --0.19317 & 0.33238 & --0.08191 \\
Mg5 & 8$j$ & --0.30518 & 0.33449 & 0.08066 \\
Mg6 & 8$j$ & --0.41574 & 0.33349 & --0.24823 \\
Mg7 & 8$j$ & --0.41491 & 0.16980 & --0.24364 \\
Mg8 & 8$j$ & --0.16952 & 0.41517 & --0.24428 \\
Mg9 & 8$j$ & --0.33085 & 0.25269 & 0.24884 \\
Mg10 & 8$j$ & --0.34067 & 0.41496 & 0.24339 \\
Mg11 & 8$j$ & --0.30906 & 0.32720 & --0.41447 \\
Mg12 & 8$j$ & 0.45712 & 0.41759 & --0.41427 \\
Mg13 & 8$j$ & 0.43813 & 0.25530 & --0.41430 \\
Mg14 & 4$i$ & 0.30910 & 0 & --0.08116 \\
Mg15 & 4$i$ & 0.19823 & 0 & 0.08487 \\
Mg16 & 4$i$ & 0.08925 & 0 & --0.24938 \\
Y1 & 8$j$ & --0.16791 & 0.35614 & 0.42821 \\
Y2 & 4$i$ & 0.42721 & 0 & 0.28123 \\
Y3 & 4$i$  & 0.23663 & 0 & --0.42916 \\
Zn1 int& 2$d$ & 0 & 1/2 & 1/2 \\
Zn2 & 8$j$ & 0.42590 & 0.11239 & --0.38565 \\
Zn3 & 4$i$ & --0.23636 & 0 & --0.38481 \\
\Hline
\end{tabular}
\end{table}

\begin{table}[htpb]
\setlength\abovecaptionskip{-3pt}
\caption{
Optimized atomic positions for 18$R$ Mg$_{58}$Zn$_6$Y$_8$, where the number of Zn vacancies $x$ = 1.
The space group is $Cm$ (\#8) and the optimized lattice parameters are $a$ = 11.17, $b$ = 19.23, $c$ = 16.04 \AA, $\beta$ = 103.05$^\circ$. 
}
\label{Coordx1}
\scriptsize
\vspace{-2mm}
\begin{tabular}{lccccc}
\Hline
Label & Wyckoff position & $x$ & $y$ & $z$ \\
\Hline
Mg1  & 4$b$  & 0.44074   &  0.41757  &  0.08162   \\
Mg2  & 4$b$  & 0.05906   &  0.08202  &  $-$0.08167  \\
Mg3  & 4$b$  & 0.44451   &  0.25094  &  0.08321   \\
Mg4  & 4$b$  & 0.05348   &  0.24846  &  $-$0.08278  \\
Mg5  & 4$b$  & 0.44302   &  0.08191  &  0.08499   \\
Mg6  & 4$b$  & 0.05454   &  0.41778  &  $-$0.08460  \\
Mg7  & 4$b$  & 0.19154   &  0.33272  &  0.08155   \\
Mg8  & 4$b$  & 0.30621   &  0.16766  &  $-$0.08230  \\
Mg9  & 4$b$  & 0.30474   &  0.33448  &  $-$0.08086  \\
Mg10 & 4$b$  & 0.19320   &  0.16572  &  0.08063   \\
Mg11 & 4$b$  & 0.41179   &  0.33373  &  0.24755   \\
Mg12 & 4$b$  & 0.08570   &  0.16665  &  0.75104   \\
Mg13 & 4$b$  & 0.41134   &  0.16854  &  0.24506   \\
Mg14 & 4$b$  & 0.08723   &  0.32998  &  0.75670   \\
Mg15 & 4$b$  & 0.16412   &  0.41580  &  0.24325   \\
Mg16 & 4$b$  & 0.33169   &  0.08414  &  0.75510   \\
Mg17 & 4$b$  & 0.33272   &  0.25230  &  0.75066   \\
Mg18 & 4$b$  & 0.16600   &  0.24676  &  0.24871   \\
Mg19 & 4$b$  & 0.34313   &  0.41518  &  0.75626   \\
Mg20 & 4$b$  & 0.15436   &  0.08307  &  0.24557   \\
Mg21 & 4$b$  & 0.30750   &  0.32543  &  0.41307   \\
Mg22 & 4$b$  & 0.19353   &  0.17180  &  0.58467   \\
Mg23 & 4$b$  & 0.54112   &  0.41858  &  0.41159   \\
Mg24 & 4$b$  & $-$0.04048  &  0.08227  &  0.58493   \\
Mg25 & 4$b$  & 0.56195   &  0.25608  &  0.41446   \\
Mg26 & 4$b$  & $-$0.05953  &  0.24454  &  0.58448   \\
Mg27 & 2$a$  & 0.68999   &  0  &  0.07988   \\
Mg28 & 2$a$  & 0.30867   &  0  &  $-$0.08056  \\
Mg29 & 2$a$  & 0.80151   &  0  &  $-$0.08515  \\
Mg30 & 2$a$  & 0.19859   &  0  &  0.08819   \\
Mg31 & 2$a$  & $-$0.09597  &  0  &  0.24742   \\
Mg32 & 2$a$  & 0.09175   &  0  &  0.74952   \\
Y1   & 4$b$  & 0.17178   &  0.35505  &  0.57232   \\
Y2   & 4$b$  & 0.33416   &  0.13986  &  0.42977   \\
Y3   & 2$a$  & 0.57655   &  0  &  0.71997   \\
Y4   & 2$a$  & 0.42196   &  0  &  0.29026   \\
Y5   & 2$a$  & 0.76610   &  0  &  0.42798   \\
Y6   & 2$a$  & 0.24472   &  0  &  0.56658   \\
Zn1 int & 2$a$  & 0.50864   &  0  &  0.50393   \\
Zn2  & 4$b$  & 0.57450   &  0.11223  &  0.38668   \\
Zn3  & 4$b$  & $-$0.07059  &  0.38746  &  0.61496   \\
Zn4  & 2$a$  & 0.76500   &  0  &  0.61394   \\
\Hline
\end{tabular}
\end{table}

\begin{table}[htpb]
\setlength\abovecaptionskip{-3pt}
\caption{
Optimized atomic positions for 18$R$ Mg$_{58}$Zn$_4$Y$_8$, where the number of Zn vacancies $x$ = 3. 
The space group is $Cm$ (\#8) and the optimized lattice parameters are $a$ = 11.09, $b$ = 19.21, $c$ = 16.0 \AA, $\beta$ = 103.33$^\circ$.
}
\label{Coordx3}
\vspace{-2mm}
\begin{tabular}{lccccc}
\Hline
Label & Wyckoff position & $x$ & $y$ & $z$ \\
\Hline
Mg1 & 4$b$ & 0.44022 & 0.41795 & 0.08205  \\
Mg2 & 4$b$ & 0.05644 & 0.08253 & $-$0.08234 \\
Mg3 & 4$b$ & 0.44654 & 0.25197 & 0.08349  \\
Mg4 & 4$b$ & 0.05529 & 0.24953 & $-$0.08277 \\
Mg5 & 4$b$ & 0.44517 & 0.08257 & 0.08337  \\
Mg6 & 4$b$ & 0.05486 & 0.41816 & $-$0.08815 \\
Mg7 & 4$b$ & 0.19369 & 0.33221 & 0.08213  \\
Mg8 & 4$b$ & 0.30716 & 0.16678 & $-$0.08224 \\
Mg9 & 4$b$ & 0.30594 & 0.33414 & $-$0.08021 \\
Mg10 & 4$b$ & 0.19482 & 0.16548 & 0.08088 \\
Mg11 & 4$b$ & 0.41633 & 0.33348 & 0.25022 \\
Mg12 & 4$b$ & 0.08492 & 0.16639 & $-$0.24629 \\
Mg13 & 4$b$ & 0.41399 & 0.17034 & 0.24380  \\
Mg14 & 4$b$ & 0.07991 & 0.33263 & $-$0.24763 \\
Mg15 & 4$b$ & 0.17082 & 0.41562 & 0.24407  \\
Mg16 & 4$b$ & 0.33498 & 0.08575 & $-$0.24863 \\
Mg17 & 4$b$ & 0.32925 & 0.25525 & $-$0.24680 \\
Mg18 & 4$b$ & 0.16945 & 0.24767 & 0.25156  \\
Mg19 & 4$b$ & 0.33747 & 0.41846 & $-$0.24800 \\
Mg20 & 4$b$ & 0.15914 & 0.08563 & 0.24370  \\
Mg21 & 4$b$ & 0.30985 & 0.32795 & 0.41752  \\
Mg22 & 4$b$ & 0.19334 & 0.17516 & $-$0.41042 \\
Mg23 & 4$b$ & $-$0.45473 & 0.41733 & 0.41723  \\
Mg24 & 4$b$ & $-$0.03922 & 0.08062 & $-$0.41049 \\
Mg25 & 4$b$ & $-$0.43817 & 0.25479 & 0.41772  \\
Mg26 & 4$b$ & $-$0.06455 & 0.24427 & $-$0.41014 \\
Mg27 & 2$a$ & $-$0.30963 & 0 & 0.07941          \\
Mg28 & 2$a$ & 0.30811 & 0 & $-$0.08105          \\
Mg29 & 2$a$ & $-$0.20055 & 0 & $-$0.08888         \\
Mg30 & 2$a$ & 0.19726 & 0 & 0.08438           \\
Mg31 & 2$a$ & $-$0.08828 & 0 & 0.25163          \\
Mg32 & 2$a$ & 0.09531 & 0 & $-$0.24764          \\
Y1 & 4$b$ & 0.16069 & 0.36244 & 0.43073     \\
Y2 & 4$b$ & 0.33536 & 0.14204 & 0.43244  \\
Y3 & 2$a$ & $-$0.43404 & 0 & $-$0.30159        \\
Y4 & 2$a$ & 0.42555 & 0 & 0.27588          \\
Y5 & 2$a$ & $-$0.23881 & 0 & 0.43291         \\
Y6 & 2$a$ & 0.24781 & 0 & $-$0.43169         \\
Zn1 int & 2$a$ & 0.49618 & 0 & 0.48936          \\
Zn2 & 4$b$ & $-$0.42681 & 0.11199 & 0.38409 \\
Zn3 & 2$a$ & 0.23739 & 0 & 0.38377          \\
\Hline
\end{tabular}
\end{table}

\bibliographystyle{jpsj}
\bibliography{./Mg_LPSO}